\documentclass[12pt]{iopart} 

\usepackage{epsfig}

\newcommand{\beq}{\begin{equation}}
\newcommand{\eeq}{\end{equation}}
\newcommand{\beqa}{\begin{eqnarray}}
\newcommand{\eeqa}{\end{eqnarray}}

\def\fun#1#2{\lower3.6pt\vbox{\baselineskip0pt\lineskip.9pt
  \ialign{$\mathsurround=0pt#1\hfil##\hfil$\crcr#2\crcr\sim\crcr}}}

\begin{document} 

\title{Entanglement entropy and the simulation of Quantum Mechanics}
\author{Jos\'e I. Latorre} 
\address{Dept. d'Estructura i Constituents de la Mat\`eria, 
Universitat de Barcelona, Mart\'\i\ i Franqu\`es 1, 08028 Barcelona, Spain.} 

\date{\today} 

\begin{abstract} 

The relation between entanglement entropy and 
the computational difficulty of classically simulating Quantum
Mechanics is briefly reviewed. Matrix product states are
proven to provide an efficient representation of one-dimensional 
quantum systems. 
Further applications of the techniques based on 
matrix product states, some of their spin-off and their 
recent generalizations to scale invariant theories and higher
dimensions systems  are also discussed.
\footnote{Contribution to the Proceedings of the IRGAC conference held
at Barcelona, July 2006.} 

\end{abstract} 


\section{Entanglement entropy as a measure of quantum correlations } \label{sec:intro}

A common misconception states that, in general, large quantum mechanical system can not be 
efficiently described by classical means. This prejudice can be illustrated
with the simple example of a system composed of $n$ two-level systems or 
qubits.
The Hilbert space of this system corresponds to the direct product 
$C^{2\otimes n}$ and an arbitrary state can be expressed in
the natural (also called computational) basis
\beq
\vert\psi\rangle=\sum_{i_1,i_2,\ldots,i_n=0,1} c^{i_1i_2\ldots i_n}
\vert i_1,i_2,\ldots,i_n\rangle .
\eeq
In order to fully specify an arbitrary state, it seems 
necessary to provide all the 
$c^{i_1\ldots i_n}$ coefficients, 
that is, $2^n$ complex numbers (minus a global phase
and  a normalization constraint that we can ignore 
for the counting of the scaling of needed resources). As $n$ grows,
the classical representation of a quantum state requires
exponential resources. Furthermore,
the processing of the state, {\sl e.g.} the computation 
of its time evolution, and
the computation of observables also requires exponentially 
many operations. 

The exponential effort needed to deal with Quantum Mechanics
can also be advocated using  an  argument based on entropy. 
The precise statement says that an average random
state in the Hilbert space is known to carry maximal 
von Neumann entropy. Let us describe in more detail this point.
Consider a partition of the original state into two parties, 
$A$ and $B$. If party $A$ ignores party $B$, the description of
its subsystem is based on the reduced density matrix
\beq
\rho_A={\rm tr}_B \vert\psi\rangle\langle\psi\vert .
\eeq
The description that party $A$ is making of the system ignores
quantum correlations between $A$ and $B$. If $A$ would suddenly
discover that it was correlated to $B$ a surprise would take
place. The amount of that surprise is quantified by the
von Neumann entropy
\beq
S(\rho_A)=-{\rm tr}\left(\rho_A\log\rho_A\right) .
\eeq
It is well-known that the entropy attached to party $A$
ignoring party $B$ equals the reciprocal one, that is, the
entropy attached to party $B$ when ignoring party $A$. This is a
consequence of the Schmidt decomposition
\beq
\vert \psi\rangle=\sum_{a=1}^{\chi=\min(dim H_A, dim H_B)}
\lambda_a\vert \xi_a\rangle_A\vert \varphi_a\rangle_B 
\eeq
with real $\lambda_a\ge 0$, 
$\sum_a \lambda_a^2=1$, and $\vert \xi_a\rangle_A$ and
$\vert \varphi_a\rangle_B$ being new orthonormal  basis for  
parties $A$ and $B$.
The magic of this decomposition is that it provides
a basis such that any state  is written with a minimum number of
coefficients $\chi$, called Schmidt number, and
the corresponding changes 
$\vert\xi_a\rangle=\Gamma^i_a\vert i\rangle_A$
and $\vert\varphi_a\rangle=\Gamma'^j_a 
\vert j\rangle_B$ to the computational basis. A prominent
example is the economical description of a product
state since $\chi=1$ and only the changes of basis are
to be retained. For the same token, the Schmidt number $\chi$
can be understood as a measure of entanglement. A more
sophisticated measure of entanglement is the von Neumann entropy
which reads
\beq
S(\rho_A)=S(\rho_B)=-\sum_{a=1}^\chi \lambda_a\log\lambda_a \ .
\eeq

We can now come back to the alleged misconception
on the exponential difficulty to deal with any
quantum system. The argument says that the totally
random state made with $n$ spins,
$n=n_A+n_B$, is such that all the eigenvalues in the
reduced density matrix $\rho_A$ are identical and equal
to $2^{-n_A}$. Thus,
$S(\rho_A)= n_A$,
which is the maximum possible scaling of the entropy
for that subsystem. Thus, the entanglement entropy scales with
the volume of the subsystem which corresponds to
maximal entropy and  quantum correlations pervade
the system.

\section{Refutation of the need for exponential resources}
\label{sec:refutation}

The argument stating that the representation
of any $n$-body quantum system needs exponential (in $n$) resources does
not apply as a general rule. The reason that
invalidates the general argument is  
a combination of two facts:
\begin{itemize}
\item It is not
necessary to represent a given state 
in the original computational basis. 
This should come as no surprise since
we are used to compress information. Consider a piece of literature.
To keep all the information about the text, it is not
necessary to write all its characters. We can  define
a clever conversion table and use shorter characters for 
frequent words. This procedure can be made close to optimal for 
arbitrary and long sequences using entropic compression codes (as the 
Lempel-Ziv based {\sl gzip}\cite{LZ77}). 
We shall later argue that we already have
techniques to represent and manipulate quantum systems 
which are far better than the naive computational basis.
In this sense, we do know compressions methods for Quantum
Mechanics.

\item In general, typical physical
states are not random. Local Hamiltonians produce interactions
between neighboring particles. The quantum correlations that 
pervade the system are far less than the maximum possible.
In other words, typical physical states do not carry
maximal entropy.
\end{itemize}

In recent years, some intense research has
addressed the problem of finding an
optimal classical representation for relevant quantum
systems. Depending on the problem,
three main ideas are currently pursued. Whenever possible,
exact simulations are carried out. In practice, this is possible only
for systems of few particles as shown by the 
work done on cold gases of few particles.
A second avenue of work are Monte Carlo simulations. This is,
for instance,
the standard technique to investigate quantum field theories
regularized on a lattice.  The method allows for computations
of correlators but it is not appropriate for the detailed simulation
of time evolution of quantum systems, neither to get a good
grasp on specific wavefunctions as {\sl e.g.} the ground state.
Furthermore, the lattice approach faces the
so-called sign problem.
A third idea to represent quantum systems looks for a specific
basis where correlations are well-represented, that we shall 
now address.

\section{Matrix product sates}
\label{sec:mps}

Let's consider a $n$-particle quantum product state 
\beq
\psi=\vert \xi_1\rangle \otimes\ldots\otimes\vert\xi_n\rangle=
(\alpha_1 \vert 0\rangle+\beta_1 \vert 0\rangle)\otimes
\ldots\otimes  (\alpha_n \vert 0\rangle+\beta_n \vert 0\rangle)
\eeq
where $\alpha_i$ can be chosen real and $\vert \alpha_i\vert^2+
\vert\beta_i\vert^2=1$.
Note that this state is represented with $2n$ real numbers, far less than
the naive exponential counting of $ 2^n$ complex numbers. The reason for this saving
can be traced to the fact that all bi-partitions of the system
carry zero entropy. There is no surprise in adding
uncorrelated new particles to any subsystem.

Can this idea be pushed further? Indeed, it is possible to
find an economical basis to retain 
all the correlations in the system. The idea works in an 
iterative way. We first take the Schmidt decomposition between
the first qubit and the rest of the system. Only the change of basis
for the first qubit $\vert \alpha_1\rangle_1=\Gamma^{[1]i_1}_{\alpha_1}
\vert i_1\rangle_1$
and the $\chi_1$ eigenvalues of this decomposition
will be retained. We, then, proceed to find the Schmidt decomposition between
the first two qubits and the rest of the system. Again, we retain
$\chi_2$ eigenvalues of the decomposition and find out the change of
basis between the basis found in the first decomposition and this second
one for the second qubit that amounts to a tensor
$\vert\alpha_1\rangle_2=\Gamma^{[2]i_2}_{\alpha_1\alpha_2}
\vert \alpha_2\rangle_2$. The procedure is iterated, giving the
result \cite{Vidal}
\beq
\vert \psi \rangle=
\sum_{\alpha_1=1}^{\chi_1}
\ldots \sum_{\alpha_{n-1=1}}^{\chi_{n-1}}
\Gamma^{[1]i_1}_{\alpha_1} \lambda^{[1]}_{\alpha_1}  
\Gamma^{[2]i_2}_{\alpha_1\alpha_2} \lambda^{[2]}_{\alpha_2}\ldots  
 \lambda^{[n-1]}_{\alpha_{n-1}}
\Gamma^{[n]i_n}_{\alpha_{n-1}}  
\vert i_1,i_2,\ldots,i_n\rangle .
\eeq
This construction represents the original
coefficients $c^{i_1\ldots i_n}$ as a product of matrices,
hence the name Matrix Product State (MPS)
\cite{Fannes,mps-garching}. It is an
exact representation that is able to adapt to 
the specific entanglement content of a state.
To see this, note that a product state corresponds
to a state with $\chi_1=\ldots=\chi_{n-1}=1$, that is,
any Schmidt decomposition is made with a single term.
The more entangled a state is, the larger the matrices $\Gamma$
will be. It is possible to actually find the maximum
size of any bi-partition. Let's take a party A made of $l$
qubits versus $n-l$. Then the size of the $H_A$ 
Hilbert space is $2^l$. Thus, $\chi_l\le 2^l$. An arbitrary
state will carry maximum entropy and each matrix
will reach its maximum possible size. Yet, in most
relevant cases, the size of the matrices will be 
smaller than their maximum.

We may furthermore absorb the eigenvalues $\lambda$ into
the matrices $\Gamma$'s. We may also decide to 
extend the original MPS representation and take
an extra periodic index
and set all the matrices of equal size $\chi$. We then 
have a periodic boundary representation of the state
\beq
\vert \psi\rangle =\sum_{i_1i_2\ldots i_n=0,1}
tr\left(A^{[1]i_1}A^{[2]i_2}\ldots A^{[n]i_n}\right)
\vert i_1,i_2,\ldots,i_n\rangle .
\eeq
This expression shows the depth of the idea of matrix product
states. All coefficients $c^{i_1i_2\ldots i_n}$ are representation
as a clever multiplication of matrices. An exact representation
will need a different size for the matrices depending on the
entanglement present in the state. A simple counting shows
that the original $2^n$ coefficients are now represented with
$2 n \chi^2$ elements. It is clear that an absolute random
state will need $\chi\sim 2^{\frac{n}{2}}$. In general, though,
physical states carry less entropy and the MPS representation
becomes a powerful tool to represent them.

Let us pause for a moment and give a very simple example
that illustrates the idea underlying the compression
power of matrix product states. Let us try to communicate
a friend the set of numbers 6, 10, 15, 22, 33, 42, 63, 55, 105 and 231.
Instead of sending those ten numbers we can as well
transmit the instruction of taking all the pair multiplications
of 2, 3, 5, 11 and 21. This packing is exponentially economical if
we consider multiplications of $n$ numbers. MPS is a sophistication
of this
multiplicative saving that also handles superpositions,
that is entanglement. It is clever compression of entanglement
perfectly suited for states which are close to product states.

Let us go one step beyond and see that the size of the
matrices involved in the MPS construction is directly
related to how much entanglement that state carries.
Any partition of the system, say at site $a$,
can be viewed as a Schmidt decomposition
\beq
\psi= \sum_{\beta=1}^{\chi_a} \lambda^{[a]}_\beta 
\left(M^{i_1\ldots i_a}_{L,\beta} \vert i_1\ldots i_a\rangle\right)
\left(M^{i_{a+1}\ldots i_n}_{R,\beta} \vert i_{a+1}\ldots i_n\rangle\right)
\eeq
where $M_{L,R}$ stand for the product of matrices on the left and
on the right of the index $a$.
As a consequence, the entropy for both the left and right parties is
\beq
S(a)={\rm tr}_{a+1,\ldots,n}\vert\psi\rangle\langle\psi\vert=
\sum_{\beta=1}^{\chi_a}\lambda^{[a]}_\beta \log \lambda^{[a]}_\beta \ .
\eeq
The maximum entropy that such a state can carry corresponds to the
case where all $\lambda^{[a]}_\beta=1/\chi_a$. Then,
\beq
S(a)\le\log \chi_a\ .
\eeq
This result shows that some amount of quantum 
correlations can be described with modest values of $\chi$'s. 
It also shows that random states need exponential $\chi$'s.

It is also worth noticing that an MPS with periodic
boundary conditions will always have two indices connecting
left and right. One index works as above
 and a second one wraps around the periodic 
boundary. The argument gets modified in the sense
that $S(a)=2 \log \chi_a$ for periodic MPS, that is, periodic
MPS uses matrices with half the dimension
of the ones needed with open boundary conditions.

A final and relevant remark must be emphasized. 
The typical distributions of the eigenvalues of the 
reduced density matrix in physical
systems is not flat. In some cases, the distribution 
decays exponentially. This suggests that a truncation 
in $\chi$ may provide a sensible approximation to
the system.

\section{Entropy and matrix product states for spin chains}
\label{sec:spinchains}

We have seen that a certain amount of quantum correlations
can be described faithfully with the  MPS construction. It remains
now to know what is the amount of entanglement present
in the ground state of a typical quantum system.

This question can be fully answered for quantum spin chains.
It is possible to compute \cite{spinchains,spinchainsus,spinchainscardy}
the entropy carried by the reduced density
matrix of $l$  (out of $n\to \infty$) spins for the ground
state of a critical system 
\beq 
S_l=\frac{c}{3}\log l \ ,
\eeq
where $c$ is the central charge of the conformal field
theory that describes the universality class of the
phase transition. This amount of entanglement is far lower
than the entropy carried by 
a random state (which would be $S_l\sim l$). We can now match 
this result from our previous MPS argument to show that
the properties of this $l$-spin block are faithfully reproduced
by a periodic MPS state with size
\beq
\chi=  l^{\frac{c}{6}} \ .
\eeq
As $l$ grows, only a polynomial increase of computational effort
is needed. Thus, quantum phase transitions on spin chains can be
efficiently simulated. Indeed, the technique of 
density matrix renormalization group (DMRG) \cite{white}
has been widely
applied to one-dimensional systems with hundreds of spins. This
would definitely be impossible if the entropy would have grown
as a power of $l$ rather than a $\log l$. Yet, even the moderate
need of classical resources we have established is commonly considered
as a poor representation of critical systems. As we shall
shortly see, only non-critical theories can be described with
a precision that improves exponentially with $\chi$.

Let us note that the entropy contained in the ground state
of a spin chain corresponds to an area law \cite{arealaw,arealawus}. 
In higher dimensions,
Hamiltonians made with local interactions are expected to deliver ground
states with $S_l\sim l^{\frac{d-1}{d}}$, where $d$ stand for 
the number of spacial dimensions. For $d=1$ the power law
is substituted with a log. The area law growth of entropy must
be seen as the quantitative barrier that prevents 
faithful simulation of higher dimensional quantum systems.
Any new technique to handle quantum systems should 
aim at this problem.

As we just mentioned, 
it is also possible to compute the entropy content of spin chains
away from the quantum phase transition point. There, the entropy 
saturates to a maximum value dictated by the 
parameters of the model \cite{spinchainsus}.
An MPS approximation can then be exponentially precise. 
A large literature on the technique of the above mentioned DMRG
(which is a method to find MPS approximations to 
ground states of Hamiltonians) shows the power of 
the entropy calculation.

Further developments on the relation between entropy and renormalization
group hints at a decrease of entanglement along renormalization
group flows \cite{rg1,arealawus}. Moreover, renormalization group transformations
can be operated on states and, more specifically, on matrix product states 
\cite{rg2}. It would be very nice to obtain further results along these
lines for higher
dimensional theories.

\section{New applications on matrix product sates: 
continuous variables, Laughlin state, quantum computation}

MPS can be be used to approximate any computation
of a ground state. For instance, it is possible
to consider discretizations of quantum field theories
and work out the ground state. In reference  \cite{mpscontinuous}
it is shown how to deal with a discretized free bosonic
theory, that is a set of harmonic oscillators to
get {\sl e.g.} the entropy present in the ground state
or the eigenvalues of the reduced density matrix. 
The basic idea is to approximate the ground state
of the system with local degrees of freedom at positions
$x_1,\ldots, x_n$ with
\beq
\psi(x_1,\ldots,x_n)=
{\rm tr}(A^{[1]a_1}\ldots A^{[n]a_n}) H_{a_1}(x_1)\ldots H_{a_n}(x_n)
\eeq
where $H_a(x)$ provide a basis for the local continuous Hilbert
space ({\sl e.g.} Hermite polynomials times gaussians). Entanglement
between the basis elements is taken into account by the MPS
construction.
Furthermore, the MPS method can be extended to an infinite system
accepting that all the matrices $A$ are identical.
Then the algorithm to compute the ground state
can be made to respect translational invariance \cite{translational,mpscontinuous}. 
This variant produces MPS that right away describe the 
thermodynamical  limit of the system. Further work along
these lines is needed to assess the power
of this method.

It is also possible to approach other highly entangled systems
and represent their ground state as an MPS.
Let us consider the Laughlin wavefunction \cite{Laughlin}
\beq
\psi=A_m \prod(z_i-z_j)^m \exp{-\frac{1}{2}\sum_i \vert z_i\vert^2} \ ,
\eeq
where $\nu=1/m$ is the filling fraction in the system.
It is extremely hard to simulate this wavefunction, as 
shown by the fact that its normalization $A_m$ is unknown in 
general. If we could find an MPS realization of this
wavefunction, we could have a better chance to carry
exact computations. Let us see that for $m=1$ this
is indeed possible. Then, the wavefunction corresponds
to a fermionic system described by a Vandermonde 
determinant. The wavefunction can 
be rewritten as
\beq \psi=A' \sum_{a_1,\ldots,a_n=0}^{n-1}
\epsilon^{a_1\ldots a_n}\phi_{a_1}(z_1)
\ldots \phi_{a_n}(z_n) \ ,
\eeq
where $\epsilon$ is the Levi-Civita fully antisymmetric tensor
and $\phi_a(z)=\frac{1}{\sqrt {\pi a!}}z^a\exp{-\frac{1}{2}\vert z\vert^2}$
form a monoparticular basis. 
The way to rewrite the coefficients as a product of matrices
is simple since this is precisely a property of
the Clifford algebra \cite{mpslaughlin}
\beq 
\epsilon^{a_1\ldots a_n}=
{\rm tr}\left(\gamma^{a_1}\ldots \gamma^{a_n}\gamma_5\right)
\quad ,\quad
\{\gamma^a,\gamma^b\}=2 \delta^{ab} \qquad a,b=0,\ldots,n-1
\eeq
where $\gamma_5\equiv (-i)^{n/2} \; \gamma^0\ldots \gamma^{n-1}$
(here, we just consider even dimensions).
Note that the original wavefunction for $m=1$ would carry
an apparent number of degrees of freedom $n^n$ because
there are $n$ particles that may occupy $n$ states. An exact
computation of the entropy for half of the system shows
that $S(n/2)=\log \left({n\atop n/2}\right)\sim n$.
The periodic MPS state uses matrices whose dimension
${\rm dim}\gamma^a= 2^{[n/2]}$ exactly matches the entropy,
$\log \chi^2=n$ in the limit $n\to \infty$.
Hence, the MPS construction is optimal.
The cases with larger $m$ can be constructed by
using a direct product construction of $\gamma$ matrices.
That construction is not optimal since the
entropy for an arbitrary $m$ Laughlin wavefunction is
known to be bounded by $n\log m$ whereas the direct product
construction needs ${\cal O}(n m)$ elements.

Let us also mention that some work has pushed 
the application of MPS
to entirely new settings. It is possible, for instance,
to simulate the whole evolution of a quantum algorithm
using MPS techniques \cite{mpsnpcomplete}. The initial state is 
represented as a MPS and then a series of
non-local quantum gates are applied
as an adiabatic evolution driven by a problem Hamiltonian. 
It has been possible to solve some NP-complete problem
with up to 100 qubits. The one solution, out
of $2^{100}$ possibilities, of a hard
problem has been deterministically obtained using
an MPS simulation of a quantum algorithm.

\section{Spin-off: 
image compression, differential equations}
\label{sec:spin-off}

It is tantalizing to try to develop some spin-off 
applications of MPS beyond Quantum Mechanics.
Two ideas have already been worked out.

The first one consists of using MPS truncation techniques
to compress an image \cite{photo}. Let us start by mapping
an image into a quantum pure (real) state.
Take a telescopic addressing of pixels in quadrants
organized as follows.
A pixel lying in the first quadrant carries
a label $\vert 1\rangle$ (or $\vert 2\rangle$,
$\vert 3\rangle$ or $\vert 4\rangle$ for the other options).
Each quadrant is subdivided again in fourths.
The new labeling for a pixel in quadrant 1, sub quadrant 2,
is $\vert 12\rangle$. We can proceed up to $n$ levels, so that
the image is made by $4^n$ pixels. Each pixel carries
a grey level that we use as its coefficient. Then 
\beq
4^n {\rm pixel\ image}\rightarrow \vert \psi\rangle= \sum_{i_1\ldots i_n=1}^4
c^{i_1\ldots i_n}\vert i_1\ldots, i_n\rangle
\eeq
represents a $4^n$-pixel grey image where the basis 
spans over all pixels and the coefficient
of each basis element gives the grey level of the 
corresponding pixel.
It is trivial to turn these coefficients into 
an MPS. A truncation of the size $\chi$ of the matrices in the 
MPS is tantamount to a compression of the picture.
Results are remarkably competitive.

A second idea to use MPS outside the domain of
Quantum Mechanics is to solve partial differential equations
\cite{mpscontinuous}.
A partial differential equation with $n$ variables can be viewed
as an operator acting on the variables and coupling them.
This is just another form of entanglement.
We can take the operator in the differential equation and
turn it into a continuous variable problem that
can be addressed using the continuous variable
techniques presented
in the previous section. A minimum distance principle
emerges as the error in the solution of
the equation. Again, the results obtained are surprisingly
good and deserve further attention.

\section{Beyond MPS: MERA and PEPs}

The shortcoming of MPS is the limited amount of entanglement they
can support. Let us take the ground state of a Hamiltonian
with local interactions defined on a quantum network in two dimensions.
We expect that any geometrical partition of this state will carry 
an area law entanglement, that is, the entropy will grow linearly
as the  number of degrees of freedom that define the boundary
of the chosen partition. 
Therefore, there is no good representation of the ground state 
in terms of MPS as $\chi$ should grow exponentially.
This is the reason why there are no faithful simulations of higher
dimensional quantum systems. In other words, we need a new technique
that beats the  area law scaling of entanglement.

Two ideas have been launched in recent years to overcome
MPS shortcomings. The first one carries the name of 
Multiscale Entanglement Renormalization Ansatz (MERA)
\cite{mera} and proposes a new
way to organize the book-keeping of entanglement using 
renormalization group ideas to improve on MPS.
MERA are built so as to represent quantum systems at
a critical point. They combine the block-spin idea with
a set of disentangling operations that optimize
the way entanglement is retained and manipulated. 

A second idea is directly constructed to deal with higher dimensional
systems. It extends the matrix product idea to a tensor contraction.
This new tensor representation carries the name of Projected
Entangled Pairs (PEPs) \cite{PEPS}. 
PEPs are proven to support area law entanglement. 
It is also known that the physical construction of PEPs is equivalent 
to solving NP-complete problems. An algorithm to find the PEPs that
describe the ground state of a quantum network is already available.

The conclusion of recent research remains open. We still don't know
what is the optimal way to represent quantum systems. Entropy computations
are no longer academic results since they establish the amount
of entanglement to be represented. MPS are proven efficient on
one-dimensional systems. A lot of work is still needed on
critical systems and higher dimensions to have fully satisfactory 
answers to delimit the classical resources necessary to faithfully
represent and manipulate quantum mechanical states. 

\ack 
I am in real debt with present and past collaborators 
on this subject from the University of
Barcelona: S. Iblisdir, R. Or\'us, E. Rico, A. Riera and G. Vidal. 
This work was supported by grants form QAP (EU), MEC and Generalitat 
de Catalunya.

\end{document}